\documentclass[prd,aps,twocolumn,preprintnumbers, showpacs, nofootinbib,superscriptaddress,notitlepage]{revtex4-2}
\usepackage{amssymb,amsthm,amsmath}
\usepackage{textcomp}
\usepackage{subfig,graphicx}   
\usepackage{color}      
\usepackage{slashed}    
\usepackage{verbatim}
\usepackage[normalem]{ulem}
\usepackage{rotating}   
\usepackage{multirow}  

\usepackage[colorlinks,
linkcolor=red,
anchorcolor=blue,
citecolor=blue
]{hyperref}

\begin{document}

	\baselineskip=15pt
	
	\preprint{CTPU-PTC-25-41}

\title{Predictions for CP violation in anti-triplet beauty baryon to charmonium decays}

\affiliation{Particle Theory and Cosmology Group, Center for Theoretical Physics of the Universe, Institute for Basic Science (IBS), Daejeon 34126, Korea }
\affiliation{ Department of Physics and Institute of Theoretical Physics, Nanjing Normal University, Nanjing, Jiangsu 210023, China}
\affiliation{State Key Laboratory of Dark Matter Physics, Tsung-Dao Lee Institute and School of
Physics and Astronomy, Shanghai Jiao Tong University, Shanghai 201210, China}
\author{Jin Sun$^{1}$}
\email{sunjin0810@ibs.re.kr}
\author{Zhi-Peng  Xing$^{2}$}
\email{zpxing@nnu.edu.cn(corresponding author)} 
\author{Xiao-Gang He$^{3}$}
\email{hexg@sjtu.edu.cn}

\begin{abstract}

Motivated by the recent 3.9$\sigma$ evidence for CP violation from the LHCb collaboration in decays of an anti-triplet beauty baryon to a charmonium, an octet baryon, and a pseudoscalar meson, 
we perform, for the first time, a systematic analysis of this class of decays within the framework of flavor $SU(3)$ symmetry. 
 Several predictions for branching ratios and CP violating relations which can be tested in future experiments are found, in particular $\Delta A_{CP} = A_{CP}(\Lambda_b^0 \to  p \pi^- J/\psi) - A_{CP}(\Lambda_b^0 \to  p K^- J/\psi) = A_{CP}(\Lambda_b^0 \to  n \pi^0 J/\psi) - A_{CP}(\Lambda_b^0 \to (n K_S, n K_L) J/\psi)$. Our results provide guidance to test the availability of $SU(3)$ symmetry and search for possible CP violation effects in this class of decays.

\end{abstract}

\maketitle

\noindent{\bf Introduction}
\\

Charge-parity (CP) violation is a cornerstone of modern particle physics, which is a crucial ingredients for generating the observed matter-antimatter asymmetry in the universe~\cite{Sakharov:1967dj}.
Within the Standard Model (SM), CP violation is incorporated via the Cabibbo-Kobayashi-Maskawa (CKM) mechanism, which describes quark mixing across three generations.
Consequently, the study of CP violation serves to both probe the origins of matter-antimatter asymmetry and stringently test the SM. Great efforts have been made to search for evidences of CP violation in different decays~\cite{Christenson:1964fg,PDG}.


Recently, LHCb collaboration has reported evidence of CP violation in beauty-baryon decays to charmonium final states~\cite{LHCb:2025svy}.
A combination with a previous LHCb analysis~\cite{LHCb:2014nhe} yields the difference in CP asymmetries between $\Lambda_b^0\to p\pi^- J/\psi$ and $\Lambda_b^0\to pK^- J/\psi$,
\begin{eqnarray}\label{eq:CPdata}
   &&\Delta A_{CP} = A_{CP}^{dir}(\Lambda_b\to p\pi^-J/\psi)-A_{CP}^{dir}(\Lambda_b\to p K^-J/\psi)\nonumber\\
    &&\qquad\quad=(4.31\pm1.06\pm0.28)\%\;,
\end{eqnarray}
where the first uncertainty is statistical and the second is systematic. The combined result corresponds to a significance level of $3.9\sigma$.

Although experimental evidence of CP violation exists in these beauty-baryon decays to charmonium final states,
 the theoretical analysis on these processes is extremely lacking.
Inspired by this results, 
we carry out a  first 
theoretical study based on the flavor $SU(3)$ symmetry of the SM. 
This study provides new predictions to guide future experimental searches for new CP-violating phenomena and to test the SM.

CP violation arises from the interference between tree-level and penguin-level amplitudes.
Predicting CP violation requires estimating the relative strong phase between these amplitudes, which is very difficult to obtain from first-principles calculations.
Various approaches have been developed to provide necessary information based on symmetry principles~\cite{Xing:2023dni,Zhong:2024qqs,Xing:2024nvg,Wang:2025bdl}.
In the current stage, flavor $SU(3)$ symmetry analysis represents 
a powerful tool for predicting CP violation~\cite{Grossman:2018ptn,Roy:2019cky,Roy:2020nyx,Bhattacharya:2022akr,Schacht:2022kuj,Bause:2022jes,Wang:2024rwf,Sun:2024mmk,He:2024unv,Iguro:2024uuw,Yang:2025orn,Chen:2025puj,He:2025msg}, 
particularly in the absence of extensive data, as it can readily yield testable predictions to guide experimental searches.

In this work we focus on the $SU(3)$ flavor analysis of the processes observed by LHCb~\cite{LHCb:2025svy}, which involve one of the beauty-baryon anti-triplet $B_{\mathbf{b\bar{3}}}$, the $\Lambda^0_b$,  into a light octet baryon $T_8$, a pseudoscalar meson $M$ and a charmonium $J/\psi$, $B_{\mathbf{\bar{3}}} \to T_8 M + J/\psi$, a class of decays that has not been extensively studied.
By  analyzing in a systematic way for this class of decays,  we find several predictions for branching ratios and CP violating relations which can be tested in future experiments, in particular $\Delta A_{CP} = A_{CP}(\Lambda_b^0 \to  p \pi^- J/\psi) - A_{CP}(\Lambda_b^0 \to  p K^- J/\psi) = A_{CP}(\Lambda_b^0 \to  n \pi^0 J/\psi) - A_{CP}(\Lambda_b^0 \to (n K_S, n K_L) J/\psi)$. Our results provide guidance to test the availability of flavor $SU(3)$ symmetry and search for possible CP violation effects in this class of decays.

\noindent{\bf The decay amplitudes of $SU(3)$ flavor symmetry}
\\

The leading effective Hamiltonian in the SM that governs the weak transition of $B_{b\bar 3} \to T_8 M J/\psi$ is given by
\begin{eqnarray}
\mathcal{H}_{eff}&=&\frac{G_F}{\sqrt{2}}\bigg(\sum_{i=1,2}C_i \lambda^c_q O_i-\sum_{j=3}^6 C_j \lambda^t_q O_j\bigg)+h.c.,
\end{eqnarray}
with the superscript $q = d(s)$ for $\Delta S = 0 (1)$ decay modes. Here we adopt the notations $\lambda^c_q=V_{cb}V_{cq}^*$  
and $\lambda^t_q=V_{tb}V_{tq}^*$. 
The current-current operators $O_{1,2}$ and QCD penguin operators $O_{3\sim 6}$ are expressed by
\begin{eqnarray}
 O_1&=&(\bar{c}_{\alpha} b_\beta)_{V-A}(\bar{c}_\beta {q}_{\alpha})_{V-A},\notag\\
 O_2&=&(\bar{c}_{\alpha} b_\alpha)_{V-A}(\bar{c}_\beta {q}_{\beta})_{V-A},\notag\\
 O_{3,5}&=&\sum_{q^{\prime}=u,d,s}\left(\bar{q}_\alpha b_\alpha\right)_{V-A}\left(\bar{q}_\beta^\prime q_\beta^\prime\right)_{V\mp A},\nonumber\\ O_{4,6}&=&\sum_{q^{\prime}=u,d,s}\left(\bar{q}_\alpha b_\beta\right)_{V-A}\left(\bar{q}_\beta^\prime q_\alpha^\prime\right)_{V\mp A},
 \end{eqnarray}
where $V(A)$ denotes vector (axial-vector) coupling.

Under $SU(3)$ symmetry, the effective Hamiltonian $\mathcal{H}_{eff}$ transforms under the fundamental representation. Its non-zero matrix elements between states can be written as
\begin{eqnarray}
    H_3^T=(0,\lambda^c_d, \lambda^c_s)\;,\quad
    H_3^P=(0,\lambda^t_d, \lambda^t_s)\;,
\end{eqnarray}
where the the indices $T$ and $P$ denote the contributions from tree-level and penguin diagrams, respectively. 
In contrast to the $SU(3)$ analysis of the effective Hamiltonian for $B$ meson decays into a pair of octet pseudoscalar where there are $\mathbf{15}$, $\bar{\mathbf{6}}$ and $\mathbf{3}$~\cite{He:2018php},
in our current case the structure involving only $\mathbf{3}$ leads to a particularly simple $SU(3)$ analysis.

Before presenting the decay amplitudes for a systematic analysis, we first define the group theory representations of the involved particles in the following.
We have
\begin{eqnarray}
B_{b\bar3}&=&
\begin{pmatrix}
0& \Lambda_b^0 &\Xi_b^0 \\
-\Lambda_b^0 & 0&\Xi_b^-\\
-\Xi_b^0& -\Xi_b^-&0
\end{pmatrix},\notag\\
M_8&=&
\begin{pmatrix}
\frac{\pi^0}{\sqrt{2}}+\frac{\eta_8}{\sqrt{6}}& \pi^+ &K^+ \\
\pi^- & -\frac{\pi^0}{\sqrt{2}}+\frac{\eta_8}{\sqrt{6}}&K^0\\
K^-& \bar{K}^0&-2\frac{\eta_8}{\sqrt{6}}\end{pmatrix},\notag\\
T_8&=&
\begin{pmatrix}
\frac{\Sigma^0}{\sqrt{2}}+\frac{\Lambda^0}{\sqrt{6}}& \Sigma^+ &p \\
\Sigma^- & -\frac{\Sigma^0}{\sqrt{2}}+\frac{\Lambda^0}{\sqrt{6}}&n\\
\Xi^-& \Xi^0&-\frac{2\Lambda^0}{\sqrt{6}}
\end{pmatrix}.
\end{eqnarray}
Additionally, a singlet $\eta_1$ is also considered to obtain 
the physical $\eta$ and $\eta'$, which are
 usually expressed as  
$\eta=\cos\theta \eta_8-\sin\theta \eta_1$, 
$\eta^\prime=\sin\theta \eta_8+ \cos\theta \eta_1$.
The mixing angle $\theta$ is 
$(-14.1 \pm 2.8)^\circ$~\cite{PDG,Christ:2010dd}.



A single-index representation for $B_{b\bar 3}$ can be obtained with the anti-symmetric tensor,
 $(B_3)_ i=\epsilon_{ijk}B_{b\bar 3}^{ij}$, 
\begin{equation}
    (B_3)_i=(\Xi^-_b,\; -\Xi_b^0, \;\Lambda_b^0 )\;.
\end{equation}
Furthermore, the charmonium state $J/\psi$ is an $SU(3)$ singlet.

The total decay amplitudes receive contributions from both tree-level (${\cal T}$) and penguin (${\cal P}$) operators.
The total amplitude can thus be expressed as
\begin{eqnarray}\label{eq:MM}
    {\cal M} = \lambda^c_q{\cal M}_T +\lambda^t_q{\cal M}_P\;,\quad  q=(d,s)\;\;,
\end{eqnarray}
where the $SU(3)$ invariant  decay amplitudes ${\cal M}_{T,P}$ can be obtained 
by contracting the corresponding index as
\begin{eqnarray}
   {\cal M}_{T,P}&=&
   a_{T,P} (B_3)_i (H^{T,P}_3)^i (\overline{T_8})^j_k (M_8)^k_j\nonumber\\
   &+&
   b_{T,P}  (B_3)_i (H^{T,P}_3)^j (\overline{T_8})^i_k (M_8)^k_j\notag\\
   &+&c_{T,P} (B_3)_i (H^{T,P}_3)^j (\overline{T_8})^k_j (M_8)^i_k\nonumber\\
  &+&  d_{T,P} (B_3)_i (H^{T,P}_3)^j (\overline{T_8})^i_j \eta_1\;.\label{am}
\end{eqnarray}

\begin{table*}[htbp!]
\caption{ $SU(3)$ amplitudes in Eq.\eqref{am} for different physical processes. Here $\lambda^c_q=V_{cb}V_{cq}^*$  
and $\lambda^t_q=V_{tb}V_{tq}^*$. }\label{tab:SU(3)}\begin{tabular}{|c|c|c|c|c|c|c|c}\hline\hline
channel & $\Delta S = 0$\; amplitudes & channel & $\Delta S = 1$\; amplitudes \\\hline
$\Lambda^{0}_{b}\to p  \pi^{-} J/\Psi$ & $ b_T\lambda^c_d + b_P\lambda^t_d$& $\Xi^{0}_{b}\to \Sigma^{+}  K^{-} J/\Psi$ & $ -(b_T\lambda^c_s + b_P\lambda^t_s) $\\\hline
$\Xi^{-}_{b}\to \Xi^{-}  K^{0} J/\Psi$ & $ b_T\lambda^c_d+ b_P\lambda^t_d$ &$\Xi^{-}_{b}\to \Sigma^{-}  \bar{K}^{0} J/\Psi$ & $ b_T\lambda^c_s +  b_P\lambda^t_s $ \\\hline
$\Lambda^{0}_{b}\to n  \pi^{0} J/\Psi$ & $ -\frac{b_T\lambda^c_d + b_P\lambda^t_d}{\sqrt{2}}$&$\Xi^{0}_{b}\to \Sigma^{0}  \bar{K}^{0} J/\Psi$ & $ \frac{(b_T\lambda^c_s + b_P\lambda^t_s)}{\sqrt{2}}$\\\hline
&& $\Xi^{-}_{b}\to \Sigma^{0}  K^{-} J/\Psi$ & $ \frac{b_T\lambda^c_s+b_P\lambda^t_s  }{\sqrt{2}}$\\\hline
$\Xi^{0}_{b}\to \Sigma^{+}  \pi^{-} J/\Psi$ & $ -((a_T+b_T) \lambda^c_d+ (a_P+b_P) \lambda^t_d) $ & $\Lambda^{0}_{b}\to p  K^{-} J/\Psi$ & $ (a_T+b_T) \lambda^c_s+ (a_P+b_P) \lambda^t_s$  \\\hline
$\Xi^{0}_{b}\to \Xi^{0}  K^{0} J/\Psi$ & $ -((a_T+b_T)\lambda^c_d + (a_P+b_P)\lambda^t_d) $ & $\Lambda^{0}_{b}\to n  \bar{K}^{0} J/\Psi$ & $ (a_T+b_T)\lambda^c_s + (a_P+b_P)\lambda^t_s$\\\hline
$\Xi^{0}_{b}\to \Sigma^{-}  \pi^{+} J/\Psi$ & $ -((a_T+c_T)\lambda^c_d+  (a_P+c_P)\lambda^t_d) $ &$\Lambda^{0}_{b}\to \Xi^{-}  K^{+} J/\Psi$ & $ (a_T+c_T)\lambda^c_s+  (a_P+c_P)\lambda^t_s$\\\hline
 $\Xi^{0}_{b}\to n  \bar{K}^{0} J/\Psi$ & $ -((a_T+c_T)\lambda^c_d+(a_P+c_P)\lambda^t_d)$ &$\Lambda^{0}_{b}\to \Xi^{0}  K^{0} J/\Psi$ & $ (a_T+c_T)\lambda^c_s+(a_P+c_P)\lambda^t_s$\\\hline
 $\Xi^{0}_{b}\to p  K^{-} J/\Psi$ & $ -(a_T \lambda_d^c+a_P \lambda_d^t)$ &$\Lambda^{0}_{b}\to \Sigma^{+}  \pi^{-} J/\Psi$ & $ a_T \lambda_s^c+a_P \lambda_s^t$ \\\hline
$\Xi^{0}_{b}\to \Xi^{-}  K^{+} J/\Psi$ & $ -(a_T \lambda^c_d+ a_P \lambda^t_d) $ &$\Lambda^{0}_{b}\to \Sigma^{-}  \pi^{+} J/\Psi$ & $ a_T \lambda^c_s+ a_P \lambda^t_s$ \\\hline
&& $\Lambda^{0}_{b}\to \Sigma^{0}  \pi^{0} J/\Psi$ & $ a_T\lambda^c_s+ a_P\lambda^t_s $\\\hline
$\Lambda^{0}_{b}\to \Sigma^{-}  K^{+} J/\Psi$ & $ c_T\lambda^c_d+ c_P\lambda^t_d$ & $\Xi^{0}_{b}\to \Xi^{-}  \pi^{+} J/\Psi$ & $ -(c_T\lambda^c_s+ c_P\lambda^t_s) $\\\hline
$\Xi^{-}_{b}\to n  K^{-} J/\Psi$ & $ c_T\lambda^c_d + c_P\lambda^t_d$&$\Xi^{-}_{b}\to \Xi^{0}  \pi^{-} J/\Psi$ & $ c_T\lambda^c_s + c_P\lambda^t_s $\\\hline
$\Lambda^{0}_{b}\to \Sigma^{0}  K^{0} J/\Psi$ & $ -\frac{c_T\lambda^c_d+ c_P\lambda^t_d }{\sqrt{2}}$& $\Xi^{0}_{b}\to \Xi^{0}  \pi^{0} J/\Psi$ & $ \frac{c_T\lambda^c_s+ c_P\lambda^t_s}{\sqrt{2}}$\\\hline
$\Xi^{-}_{b}\to \Sigma^{-}  \pi^{0} J/\Psi$ & $ \frac{(c_T-b_T) \lambda^c_d+(c_P-b_P) \lambda^t_d}{\sqrt{2}}$& $\Xi^{-}_{b}\to \Xi^{-}  \pi^{0} J/\Psi$ & $ \frac{c_T\lambda^c_s + c_P\lambda^t_s}{\sqrt{2}}$\\\hline
$\Lambda^{0}_{b}\to \Lambda^{0}  K^{0} J/\Psi$ & $ \frac{(c_T-2 b_T) \lambda^c_d+(c_P-2 b_P) \lambda^t_d}{\sqrt{6}}$& $\Xi^{0}_{b}\to \Lambda^{0}  \bar{K}^{0} J/\Psi$ & $ -\frac{(b_T-2 c_T) \lambda^c_s+(b_P-2 c_P) \lambda^t_s}{\sqrt{6}}$\\\hline
$\Xi^{-}_{b}\to \Lambda^{0}  \pi^{-} J/\Psi$ & $ \frac{(b_T+c_T)\lambda^c_d +(b_P+c_P)\lambda^t_d}{\sqrt{6}}$&
$\Xi^{-}_{b}\to \Lambda^{0}  K^{-} J/\Psi$ & $ \frac{(b_T-2 c_T)\lambda^c_s + (b_P-2 c_P)\lambda^t_s}{\sqrt{6}}$\\\hline
$\Xi^{0}_{b}\to \Lambda^{0}  \pi^{0} J/\Psi$ & $ \frac{(b_T+c_T)\lambda^c_d+ (b_P+c_P)\lambda^t_d }{2 \sqrt{3}}$ & &\\\hline
$\Xi^{-}_{b}\to \Sigma^{0}  \pi^{-} J/\Psi$ & $ \frac{(b_T-c_T)\lambda^c_d +(b_P-c_P)\lambda^t_d}{\sqrt{2}}$ && \\\hline
$\Xi^{0}_{b}\to \Sigma^{0}  \pi^{0} J/\Psi$ & $ -\frac{1}{2} \lambda^c_d (2 a_T+b_T+c_T)-\frac{1}{2} \lambda^t_d (2 a_P+b_P+c_P)$ & &\\\hline\hline
$\Xi^{0}_{b}\to \Sigma^{0}  \eta J/\Psi$ & $ \frac{[\sqrt{3}c_\theta(b_T+c_T)-3\sqrt{2} s_\theta d_T]\lambda^c_d }{6}$&&\\
&$+\frac{[\sqrt{3}c_\theta(b_P+c_P)-3\sqrt{2} s_\theta d_P]\lambda^t_d }{6}$&&\\\hline
$\Xi^{-}_{b}\to \Sigma^{-}  \eta J/\Psi$ & $ \frac{[\sqrt{6}c_\theta( b_T + c_T )-6 s_\theta d_T ]\lambda^c_d}{6}$&$\Xi^{-}_{b}\to \Xi^{-}  \eta J/\Psi$ & $ -\frac{[\sqrt{6}c_\theta( 2b_T -c_T)+6 s_\theta (d_T) ]\lambda^c_s}{6}$\\
&+$\frac{[\sqrt{6}c_\theta( b_P + c_P )-6 s_\theta d_P ]\lambda^t_d}{6}$ & & $  -\frac{[\sqrt{6}c_\theta( 2b_P -c_P)+6 s_\theta (d_P) ]\lambda^t_s}{6}$ \\\hline
$\Lambda^{0}_{b}\to n  \eta J/\Psi$ & $+ \frac{[\sqrt{6}c_\theta( b_T -2  c_T)-6s_\theta(  d_T ) ]\lambda^c_d}{6}$&
$\Xi^{0}_{b}\to \Xi^{0}  \eta J/\Psi$ & $ \frac{[\sqrt{6}c_\theta( 2b_T -c_T )+6s_\theta(  d_T)]\lambda^c_s }{6}$\\
&$+ \frac{[\sqrt{6}c_\theta( b_P -2  c_P)-6s_\theta(  d_P ) ]\lambda^t_d}{6}$&&$+\frac{[\sqrt{6}c_\theta( 2b_P -c_P )+6s_\theta(  d_P)]\lambda^t_s }{6}$ 
\\\hline
$\Xi^{0}_{b}\to \Lambda^{0}  \eta J/\Psi$ & $ -\frac{[c_\theta(6a_T +b_T+c_T)-\sqrt{6}s_\theta( d_T )] \lambda^c_d}{6}$&
$\Lambda^{0}_{b}\to \Lambda^{0}  \eta J/\Psi $ & $ \frac{[c_\theta(3a_T +2b_T+2c_T)+\sqrt{6}s_\theta( d_T )] \lambda^c_s}{3}$\\
&$ -\frac{[c_\theta(6a_P +b_P+c_P)-\sqrt{6}s_\theta( d_P )] \lambda^t_d}{6}$&
&$+ \frac{[c_\theta(3a_P +2b_P+2c_P)+\sqrt{6}s_\theta( d_P )] \lambda^t_s}{3}$
\\\hline
$\Xi^{0}_{b}\to \Sigma^{0}  \eta^\prime J/\Psi$ & $ \frac{[\sqrt{3}s_\theta(b_T+c_T)+3\sqrt{2} c_\theta d_T]\lambda^c_d }{6}$ &&\\
&$+ \frac{[\sqrt{3}s_\theta(b_P+c_P)+3\sqrt{2} c_\theta d_P]\lambda^t_d }{6}$&&\\\hline
$\Xi^{-}_{b}\to \Sigma^{-}  \eta^\prime J/\Psi$ & $\frac{[6c_\theta d_T + \sqrt{6}s_\theta( b_T + c_T ) ]\lambda^c_d}{6}$ &$\Xi^{-}_{b}\to \Xi^{-}  \eta^\prime J/\Psi$ & $ -\frac{[-6c_\theta(d_T)+\sqrt{6}s_\theta( 2b_T-c_T  )]\lambda^c_s}{6}$\\
& +$\frac{[6c_\theta d_T + \sqrt{6}s_\theta( b_T + c_T ) ]\lambda^t_d}{6}$&&$ -\frac{[-6c_\theta(d_P)+\sqrt{6}s_\theta( 2b_P-c_P )]\lambda^t_s}{6}$\\\hline
$\Lambda^{0}_{b}\to n  \eta^\prime J/\Psi$ & $ \frac{[\sqrt{6}s_\theta( b_T -2  c_T)+6c_\theta(d_T ) ]\lambda^c_d}{6}$
&
$\Xi^{0}_{b}\to \Xi^{0}  \eta^\prime J/\Psi$ & $  \frac{[\sqrt{6}s_\theta( 2b_T -c_T)-6c_\theta(  d_T)]\lambda^c_s }{6}$\\
&$+ \frac{[\sqrt{6}s_\theta( b_P -2  c_P)+6c_\theta(d_P ) ]\lambda^t_d}{6}$&&
$+\frac{[\sqrt{6}s_\theta( 2b_P -c_P)-6c_\theta(  d_P)]\lambda^t_s }{6}$\\\hline
$\Xi^{0}_{b}\to \Lambda^{0}  \eta^\prime J/\Psi$ & 
$-\frac{[s_\theta(6 a_T +b_T +c_T )+\sqrt{6}c_\theta( d_T )] \lambda^c_d}{6}$&
$\Lambda^{0}_{b}\to \Lambda^{0}  \eta^\prime J/\Psi$ & $ \frac{[-\sqrt{6}c_\theta(d_T)+s_\theta( 3a_T +2b_T+2c_T)] \lambda^c_s}{3}$
\\
&$-\frac{[s_\theta(6 a_P +b_P +c_P )+\sqrt{6}c_\theta( d_P )] \lambda^t_d}{6}$
&&$ +\frac{[-\sqrt{6}c_\theta(d_P)+s_\theta( 3a_P +2b_P+2c_P)] \lambda^t_s}{3}$\\\hline
\hline
\end{tabular}
\end{table*}

Note that the coefficients  $d_{T,P}$ are only associated with decays with $\eta_1$ in the final states.
The relevant amplitudes are given in Table.~\ref{tab:SU(3)}.
There are only four invariant $SU(3)$ decay amplitudes for each of the tree and penguin amplitudes. This makes the analysis simple,  
while still yielding a lot  of testable predictions.
\\

\noindent{\bf $SU(3)$ branching ratio relations and CP violation}
\\

We begin by discussing the relations among different decay branching ratios. Several decay branching ratios have already been measured, as listed below~\cite{PDG,LHCb:2025lhk}
\begin{eqnarray}\label{eq:data}
 &&   \mathcal{B}(\Lambda_b\to J/\psi p\pi^-)=(2.6^{+0.5}_{-0.4})\times 10^{-5},\notag\\
 && \mathcal{B}(\Lambda_b\to J/\psi pK^-)=(3.2^{+0.6}_{-0.5})\times 10^{-4}.\nonumber\\
 &&\mathcal{B}(\Lambda_b\to J/\psi\Xi^-K^+)=(3.93\pm1.41)\times10^{-6},\nonumber\\
&&\mathcal{B}(\Xi_b^0\to J/\psi\Xi^-\pi^+)=(6.42\pm2.98)\times10^{-5}.
\end{eqnarray}
The masses and lifetimes for the anti-triplet beauty-baryons $B_{b\bar 3}$ are~\cite{PDG}
\begin{eqnarray}
&& m_{\Lambda_b^0}=5.62\text{GeV},\;m_{\Xi^-_b}=5.797\text{GeV},\; m_{\Xi^0_b}=5.792\text{GeV}
\notag\\
    &&\tau_{\Lambda_b}=1.468\times 10^{-12}\text{s},\;
    \tau_{\Xi_b^-}=1.570\times 10^{-12}\text{s},\;\nonumber\\
     &&\tau_{\Xi_b^0}=1.477\times 10^{-12}\text{s},
\end{eqnarray}


For the aforementioned decay processes with measured branching ratios, 
we can use isospin symmetry obtain 
 several  relations between decay amplitudes for $\Delta S=0$ processes, such as
\begin{eqnarray}\label{eq:amp1}
&& { \cal M}(\Lambda_b^0\to p \pi^- J/\psi)
= { \cal M}(\Xi_b^-\to \Xi^- K^0 J/\psi)\notag\\
&&=-\sqrt{2}{ \cal M}(\Lambda_b^0\to n\pi^0J/\psi)\;,
\end{eqnarray}
and $\Delta S=1$ processes as
\begin{eqnarray}\label{eq:amp2}
&&
{ \cal M}(\Lambda_b^0\to p K^- J/\psi)={ \cal M}(\Lambda_b^0\to n\bar K^0 J/\psi)\;,\\
&&{ \cal M}(\Lambda_b^0\to \Xi^-K^+J/\psi)={ \cal M}(\Lambda_b^0\to \Xi^0  K^0 J/\psi)\;,\nonumber\\
&&{ \cal M}(\Xi_b^0\to \Xi^- \pi^+ J/\psi)=-\sqrt{2}{ \cal M}(\Xi_b^0\to \Xi^0\pi^0  J/\psi)\nonumber\\
&&=-{ \cal M}(\Xi_b^-\to \Xi^0\pi^-J/\psi)=-\sqrt{2}{ \cal M}(\Xi_b^-\to \Xi^-\pi^0J/\psi)\nonumber.
\end{eqnarray}
We then obtain the following predictions for the  decay branching ratios numerically
\begin{eqnarray}
 && \mathcal{B}(\Xi_b^-\to \Xi^-K^0J/\psi)=(2.70^{+0.52}_{-0.41})\times 10^{-5}\;, \notag\\ 
 && \mathcal{B}(\Lambda_b^0\to n\pi^0J/\psi)=(1.3^{+0.25}_{-0.2})\times 10^{-5}\;,\notag\\
 && \mathcal{B}(\Lambda_b^0\to n \bar K^0J/\psi)=(3.2^{+0.6}_{-0.5})\times 10^{-4}\;, \notag\\ 
 && \mathcal{B}(\Lambda_b^0\to \Xi^0K^0J/\psi)=(3.93\pm1.41)\times 10^{-6}\;,\notag\\
&&  \mathcal{B}(\Xi_b^-\to \Xi^0\pi^-J/\psi)=(6.82\pm3.16)\times 10^{-5}\;,\notag\\
&&\mathcal{B}(\Xi_b^0\to \Xi^0\pi^0J/\psi)=(3.21\pm 1.49)\times 10^{-5}\;,\notag\\
&&\mathcal{B}(\Xi_b^-\to \Xi^-\pi^0J/\psi)=(3.41\pm 1.58)\times 10^{-5}\;.
\end{eqnarray}
These predictions can be tested once the relevant branching ratios are measured. In the numerical calculations, we have used the actual measured particle masses.
This accounts for phase space differences, which cause the branching ratios to deviate slightly from the symmetric  relations.


There are also other relations
from isospin symmetry among the decay amplitudes which can be obtained. We have
\begin{eqnarray}\label{eq:all}
&&\Delta S=0:\notag\\
&&{ \cal M}(\Lambda_b^0\to \Sigma^-K^+J/\psi)={ \cal M}(\Xi_b^-\to n K^-  J/\psi)\nonumber\\
&&\qquad\qquad\qquad\qquad\quad\;\;=-\sqrt{2}{ \cal M}(\Lambda_b^0\to \Sigma^0 K^0  J/\psi)\;,\nonumber\\
&&{ \cal M}(\Xi_b^0\to p K^-J/\psi)={ \cal M}(\Xi_b^0\to \Xi^- K^+  J/\psi)\;,\nonumber\\
&&{ \cal M}(\Xi_b^0\to \Xi^0 K^0 J/\psi)={ \cal M}(\Xi_b^0\to \Sigma^+ \pi^-  J/\psi)\;,\nonumber\\
 &&   { \cal M}(\Xi_b^0\to \Sigma^- \pi^+ J/\psi)={ \cal M}(\Xi_b^0\to n\bar K^0  J/\psi)\;,\nonumber\\
 &&{ \cal M}(\Xi_b^0\to \Lambda^0 \pi^0 J/\psi)={ \cal M}(\Xi_b^-\to \Lambda^0\pi^-  J/\psi)/\sqrt{2}\;,\notag\\
 && { \cal M}(\Xi_b^0\to \Sigma^0 \eta^{(\prime)} J/\psi)={ \cal M}(\Xi_b^-\to \Sigma^-\eta^{(\prime)}  J/\psi)/\sqrt{2}\;.\notag\\
&&\Delta S= 1:\notag\\ 
&&    { \cal M}(\Lambda_b^0\to \Sigma^+ \pi^- J/\psi)= { \cal M}(\Lambda_b^0\to \Sigma^- \pi^+ J/\psi)\nonumber\\
  &  &\qquad\qquad\qquad\qquad\quad\;\;=
     { \cal M}(\Lambda_b^0\to \Sigma^0 \pi^0 J/\psi),\;\nonumber\\
&&-{ \cal M}(\Xi_b^0\to \Sigma^+ K^-J/\psi)=\sqrt{2}{ \cal M}(\Xi_b^0\to \Sigma^0 \bar K^0  J/\psi)\nonumber\\
&&={ \cal M}(\Xi_b^-\to \Sigma^- \bar K^0 J/\psi)=\sqrt{2}{ \cal M}(\Xi_b^-\to \Sigma^0K^-J/\psi)\;,\nonumber\\
&&{ \cal M}(\Xi_b^0\to \Lambda^0 \bar K^0 J/\psi)=-{ \cal M}(\Xi_b^-\to \Lambda^0K^-  J/\psi)\;,\nonumber\\
 && { \cal M}(\Xi_b^-\to \Xi^- \eta^{(\prime)} J/\psi)=-{ \cal M}(\Xi_b^0\to \Xi^-\eta^{(\prime)}  J/\psi)\;.
\end{eqnarray}
Although none of the branching ratios mentioned above have been measured to date, these relations can be tested with the substantial amount of data expected in the future.

In addition to the amplitude relations discussed above, numerous relations also exist for CP violation across various decay modes due to the U-spin symmetry, which is a sub-group of $SU(3)$  symmetry. For the direct  CP asymmetry $A_{CP}^{dir}$, we have
\begin{eqnarray}
A_{CP}^{dir}= \frac{\int |{\cal M}(B_b\to B MM)|^2-\int|\bar{\cal M}(\bar B_b\to \bar B \bar M\bar M)|^2}{\int|{\cal M}(B_b\to B MM)|^2+\int|\bar{\cal M}(\bar B_b\to \bar B \bar M\bar M)|^2}, 
\end{eqnarray}
where the symbol $\int$ indicates the phase space integration. 

Since the total amplitudes are the combination of tree-level and penguin in Eq.~(\ref{eq:MM}), the amplitude for the CP-conjugate process is
\begin{eqnarray}
\bar {\cal M}&=&(\lambda^c_q)^*{\cal M} _T+(\lambda^t_q)^* {\cal M}_P\;, \; q=d,s\;\;.
\end{eqnarray}
Here $q=(d,s)$ correspond to $\Delta S=0$ and 
  $\Delta S=1$ processes, respectively. 
The squared amplitude is
\begin{eqnarray}
&&|{\cal M}(B_b\to B MM)|^2=|\lambda^c_q|^2|{\cal M}_T|^2+|\lambda^t_q|^2|{\cal M}_P|^2\notag\\
&&\qquad\qquad\qquad+2{\cal R}e(\lambda^c_q{\cal M}_T \times (\lambda^t_q)^*{\cal M}_P^*),\notag\\
&&|\bar{\cal M}(\bar B_b\to \bar B \bar M \bar M)|^2=|(\lambda^c_q)^*|^2|{\cal M}_T|^2+|(\lambda^t_q)^*|^2|{\cal M}_P|^2\notag\\
&&\qquad\qquad\qquad+2{\cal R}e((\lambda^c_q)^*{\cal M}_T \times (\lambda^t_q){\cal M}_P^*).
\end{eqnarray}
Therefore, the direct CP violation can be obtained as
\begin{eqnarray}
A_{CP}^{dir}&=&\frac{-4{\cal I}m(\lambda^c_q (\lambda^t_q)^*)\int{\cal I}m({\cal M}_{T}{\cal M}^*_ P)}{\int|{\cal M}(B_b\to B MM)|^2+\int|\bar{\cal M}(\bar B_b\to \bar B \bar M\bar M)|^2},\notag\\
\end{eqnarray}
where $\lambda^c_q\lambda^{t*}_q=V_{cb}V_{cq}^*V^*_{tb}V_{tq}$ with $q=(d,s)$.
Following the CKM matrix unitarity relation  ${\cal I}m(V_{cb}V_{cs}^*V_{tb}^*V_{ts})=-{\cal I}m(V_{cb}V_{cd}^*V_{tb}^*V_{td})$, We can directly determine the CP asymmetry  of the corresponding U-spin processes. Taking the $\Lambda_b\to p \pi^- J/\psi$ as an example, we have
\begin{eqnarray}
&|{\cal M}(\Lambda_b\to p \pi^- J/\psi)|^2-|\bar{\cal M}(\bar\Lambda_b\to \bar p \pi^+ \overline{J/\psi})|^2\\
&=-(|{\cal M}(\Xi_b^0\to \Sigma^+ K^-J/\psi)|^2-|\bar{\cal M}(\bar \Xi_b^0\to \Sigma^- K^+\overline{J/\psi})|^2).\notag
\end{eqnarray}
Then the $A_{CP}^{dir}$ ratio of these channels is
\begin{eqnarray}
&&R(A_{CP}^{dir})=\frac{A_{CP}^{dir}(\Lambda_b\to p\pi^-J/\psi)}{A_{CP}^{dir}(\Xi_b^0\to \Sigma^+ K^-J/\psi)}\notag\\
&&=-\frac{\int |{\cal M}(\Xi_b^0\to \Sigma^+ K^-J/\psi)|^2+\int|\bar{\cal M}(\bar \Xi_b^0\to \Sigma^- K^+\overline{J/\psi})|^2}{\int |{\cal M}(\Lambda_b\to p\pi^-J/\psi)|^2+\int |\bar{\cal M}(\bar\Lambda_b\to \bar p \pi^+ \overline{J/\psi})|^2}\notag\\
&&=-\frac{{\cal B}(\Xi_b^0\to \Sigma^+ K^-J/\psi)\cdot \tau(\Lambda_b)}{{\cal B}(\Lambda_b\to p\pi^-J/\psi)\cdot\tau(\Xi_b^0)} ,\label{CPV}
\end{eqnarray}
where we have used $(\int|{\cal M}(\Lambda_b\to p\pi^-J/\psi)|^2+\int |\bar{\cal M}(\bar\Lambda_b\to \bar p \pi^+ \overline{J/\psi})|^2)/2 \sim {\cal B}(\Lambda_b\to p\pi^-J/\psi) $.
Therefore we obtain the following CP violation relation
\begin{eqnarray}\label{eq:UCPV}
&&\frac{A_{CP}^{dir}(\Lambda_b\to p\pi^-J/\psi) }{A_{CP}^{dir}(\Xi_b^0\to \Sigma^+ K^-J/\psi)}=-{\cal R}\left(\frac{\Xi_b^0\to \Sigma^+ K^-J/\psi}{\Lambda_b\to p\pi^-J/\psi}\right),\notag\\
&&\frac{A_{CP}^{dir}(\Lambda^0_b\to p K^-  J/\psi)}{A_{CP}^{dir}(\Xi^0_b\to \Sigma^+\pi^-  J/\psi)}=-{\cal R}\left(\frac{\Xi^0_b\to \Sigma^+\pi^- J/\psi}{\Lambda^0_b\to p K^-  J/\psi}\right),\notag\\
\end{eqnarray}
where a ratio is defined for convenience: 
\begin{eqnarray}
{\cal R}\left(\frac{B_b\to B MM}{B^\prime_b\to B^\prime M^\prime M^\prime}\right)=\frac{{\cal B}(B_b\to B MM)\cdot \tau(B_b^\prime)}{{\cal B}(B^\prime_b\to B^\prime M^\prime M^\prime)\cdot\tau(B_b)}.
\end{eqnarray}

Since LHCb has measured the difference in CP asymmetries between the two aforementioned processes, as given in Eq.~(\ref{eq:CPdata}),
we have  
\begin{eqnarray}
  &&\Delta A_{CP}= {\cal R}\left(\frac{\Xi^0_b\to \Sigma^+\pi^- J/\psi}{\Lambda^0_b\to p K^-  J/\psi}\right)A_{CP}^{dir}(\Xi^0_b\to \Sigma^+\pi^-  J/\psi)\nonumber\\
  &&\qquad-{\cal R}\left(\frac{\Xi_b^0\to \Sigma^+ K^-J/\psi}{\Lambda_b\to p\pi^-J/\psi}\right) A_{CP}^{dir}(\Xi_b^0\to \Sigma^+ K^-J/\psi).\notag\\
\end{eqnarray}

Following the same strategy outlined in Eq.~\eqref{CPV}, we derive the following U-spin-based CP-violation relations for bottom-baryon decays as
\begin{eqnarray}\label{eq:Ucpv}
&&\frac{A_{CP}^{dir}(\Lambda_b^0\to \Sigma^+\pi^-J/\psi)}{A_{CP}^{dir}(\Xi_b^0\to p K^-J/\psi)}=-{\cal R}\left(\frac{\Xi_b^0\to p K^-J/\psi}{\Lambda_b^0\to \Sigma^+\pi^-J/\psi}\right),\notag\\
&& \frac{A_{CP}^{dir}(\Lambda_b^0\to \Sigma^-\pi^+ J/\psi)}{A_{CP}^{dir}(\Xi^0_b\to\Xi^- K^+J/\psi)}=-{\cal R}\left(\frac{\Xi^0_b\to\Xi^- K^+J/\psi}{\Lambda_b^0\to \Sigma^-\pi^+ J/\psi}\right),\notag\\
&&\frac{A_{CP}^{dir}(\Lambda_b^0\to n\pi^0 J\psi)}{A_{CP}^{dir}(\Lambda_b^0\to \Sigma^0 \bar K^0J/\psi)}=-{\cal R}\left(\frac{\Xi^0_b\to \Sigma^0 \bar K^0J/\psi}{\Lambda^0_b\to n\pi^0 J\psi}\right),\notag\\
&&\frac{A_{CP}^{dir}(\Lambda^0_b\to n \bar K^0 J/\psi)}{A_{CP}^{dir}(\Xi^0_b\to n K^0 J/\psi)}=-{\cal R}\left(\frac{\Xi^0_b\to n K^0 J/\psi}{\Lambda^0_b\to n \bar K^0 J/\psi}\right),\notag\\
&&\frac{A_{CP}^{dir}(\Lambda^0_b\to \Xi^- K^+  J/\psi)}{A_{CP}^{dir}(\Xi^0_b\to \Sigma^-\pi^+  J/\psi)}=-{\cal R}\left(\frac{\Xi^0_b\to \Sigma^-\pi^+ J/\psi}{\Lambda^0_b\to \Xi^- K^+  J/\psi}\right),\notag\\
&&\frac{A_{CP}^{dir}(\Lambda^0_b\to \Xi^0 K^0  J/\psi)}{A_{CP}^{dir}(\Xi^0_b\to n\bar K^0  J/\psi)}=-{\cal R}\left(\frac{\Xi^0_b\to n\bar K^0 J/\psi}{\Lambda^0_b\to  \Xi^0 K^0  J/\psi}\right),\notag\\
&&\frac{A_{CP}^{dir}(\Lambda^0_b\to \Sigma^-K^+   J/\psi)}{A_{CP}^{dir}(\Xi^0_b\to \Xi^-\pi^+  J/\psi)}=-{\cal R}\left(\frac{\Xi^0_b\to \Xi^-\pi^+ J/\psi}{\Lambda^0_b\to  \Sigma^-K^+ J/\psi}\right),\notag\\
\end{eqnarray}
With the continuous accumulation of new experimental data, the observation of new processes is anticipated in the future.
Once future experiments measure the branching ratio of one or several of the above processes, we can predict the corresponding other processes.
Moreover, the relevant CP violation can also further predicted.\\

	Additionally, isospin symmetry combined with the U-spin symmetry also provides new information about CP asymmetry differences from the known $\Delta A_{CP}$ in Eq. (\ref{eq:CPdata}) for other decays. We give an example below.

Using the relations
\begin{eqnarray}
  &&  K_1=\frac{1}{\sqrt{2}}(K^0- \bar K^0),\; 
    K_2=\frac{1}{\sqrt{2}}(K^0+ \bar K^0),\;\\
&&  K_{S}^0=\frac{1}{\sqrt{1+|\epsilon|^2}}(K_1+\epsilon K_2),\; \nonumber\\
&&   K_{L}^0=\frac{1}{\sqrt{1+|\epsilon|^2}}(K_2+\epsilon K_1)\;,\notag
\end{eqnarray}
 and results in Table I, we obtain
\begin{eqnarray}\label{eq:CP}
 &&  \Delta A_{CP} = A_{CP}(\Lambda_b^0 \to  p \pi^- J/\psi) - A_{CP}(\Lambda_b^0 \to  p K^- J/\psi) \notag\\
  & &\quad = A_{CP}(\Lambda_b^0 \to  n \pi^0 J/\psi) - A_{CP}(\Lambda_b^0 \to n K_{S,L} J/\psi) \; \notag\\ 
  && \quad = A_{CP}(\Xi_b^-\to \Xi^-  K_{S, L} J/\psi) - A_{CP}(\Lambda_b^0 \to n K_{S,L} J/\psi) \;. \nonumber\\
\end{eqnarray}
Although measuring  the channel with neutron is currently challenging, it can become feasible with improvements in data analysis methods.

	For processes not yet measured, differences in CP asymmetries can be predicted using specific combinations of the amplitude relations in Eqs. (\ref{eq:all}).
	Some representative examples are
	\begin{eqnarray}
		&& A_{CP}(\Lambda_b^0\to \Sigma^-K^+J/\psi) - A_{CP}(\Lambda_b^0\to \Sigma^- \pi^+ J/\psi) \\%
		& &\quad = A_{CP}(\Lambda_b^0\to \Sigma^0  K^0 J/\psi) - A_{CP}(\Lambda_b^0\to \Sigma^0\pi^0  J/\psi),  \notag\\ 
		&& A_{CP}(\Xi_b^0\to \Lambda^0\pi^0J/\psi) - A_{CP}(\Xi_b^0\to \lambda K_{S,L} J/\psi) \notag\\%
		& &\quad =A_{CP}(\Xi_b^-\to \Lambda^0\pi^-J/\psi) - A_{CP}(\Xi_b^-\to \Lambda K^- J/\psi).\notag 
	\end{eqnarray}

 Since these intriguing difference of CP asymmetry relations can remove some systematic errors compared with individual CP asymmetry measurements, such relations warrant further experimental investigation.
Notably, current experiments are already measuring key observables, such as $A_{CP}(\Lambda_b^0 \to p \pi^- J/\psi) - A_{CP}(\Lambda_b^0 \to p K^- J/\psi)$ and the corresponding branching ratios.
A particularly compelling candidate to test our prediction is the asymmetry difference $A_{CP}(\Lambda_b^0 \to n \pi^0 J/\psi) - A_{CP}(\Lambda_b^0 \to (n K_S, n K_L) J/\psi)$, which benefits from having identical baryons  in the initial and final states.
\\

\noindent{\bf Conclusion}\label{sec4}
\\

In this work, we perform,  for the first time, a systematic analysis of the decays of antitriplet beauty baryons, $\mathbf{B}_{b\bar{3}} \to T_8 M J/\psi$, within the framework of flavor $SU(3)$ symmetry. 
The $SU(3)$ singlet feature of the charmonium state $J/\psi$ significantly simplifies the dynamics. Specifically, there exist only four independent $SU(3)$-invariant amplitudes for both the tree and penguin operators.
We categorize the decay processes into $\Delta S=0$ and $|\Delta S|=1$ channels, where $S$ denotes strangeness. By decomposing the relevant decay amplitudes, we derive a set of symmetry relations among them, which lead to testable predictions for the corresponding branching ratios. Furthermore, from the perspectives of U-spin and isospin symmetry, we establish numerous relations for CP-violating observables across different decay modes, as detailed in Eqs.~(\ref{eq:UCPV}, \ref{eq:Ucpv}, \ref{eq:CP}).
These results provide a clear guidance for testing flavor $SU(3)$ symmetry in the SM and  searching for new sources of CP violation in the three-body charmonium decays of beauty baryons. 
One particularly interesting  and testable prediction is the difference in direct CP asymmetries,
 $\Delta A_{CP} = A_{CP}(\Lambda_b^0 \to p \pi^- J/\psi) - A_{CP}(\Lambda_b^0 \to p K^- J/\psi) = A_{CP}(\Lambda_b^0 \to n \pi^0 J/\psi) - A_{CP}(\Lambda_b^0 \to (n K_S, n K_L) J/\psi)$.
 We strongly encourage our experimental colleagues to carry out related searches.
\\

\section*{Acknowledgments}

We thank Prof. Jibo He for useful discussions on the search for the neutron channel in the experiment.
The work of Jin Sun is supported by IBS under the project code, IBS-R018-D1. 
The work of Zhi-Peng Xing is supported by NSFC under grant   No. 12405113. 
 X.-G.H. was supported by the Fundamental Research
Funds for the Central Universities, by the National Natural Science Foundation of the People’s
Republic of China (Nos. 12090064, 11735010, 12205063, 11985149, 12375088, and W2441004),
and by MOST 109-2112-M-002-017-MY3.

\bibliographystyle{JHEP}
\bibliography{main}

\end{document}